\documentclass{llncs}

\usepackage{graphicx}

\usepackage{url}

\def\O2{$\textnormal{O}_2$}

\begin{document}

\title{Comparing Bug Finding 
       Tools with Reviews and Tests\thanks{This
       research was supported in part by the \emph{Deutsche 
       Forschungsgemeinschaft (DFG)} within the project
       \emph{InTime}.}}

\author{Stefan Wagner\inst{1} \and Jan J\"{u}rjens\inst{1} 
        \and Claudia Koller\inst{1} 
        \and Peter Trischberger\inst{2}}

\institute{Institut f\"{u}r Informatik\\ 
Technische Universit\"{a}t M\"{u}nchen\\
Boltzmannstr. 3, D-85748 Garching, Germany
\and
$\textnormal{O}_2$ Germany\\
Georg-Brauchle-Ring 23-25,
D-80992 Munich, Germany}

\maketitle

\begin{abstract}
Bug finding tools can find defects in software source code using an
automated static analysis. This automation may be able to reduce the
time spent for other testing and review activities. For this we need
to have a clear understanding of how the defects found by bug finding
tools relate to the defects found by other techniques. This paper
describes a case study using several projects mainly from an 
industrial environment
that were used to analyse the interrelationships. The main finding is
that the bug finding tools predominantly find different defects than
testing but a subset of defects found by reviews. However, the types
that can be detected are analysed more thoroughly. Therefore, a combination 
is most advisable if the high number of false positives of the tools can be tolerated.
\end{abstract}

\begin{center}
\tiny{The final publication is available at Springer via \url{https://doi.org/10.1007/11430230_4}.}
\end{center}

\section{Introduction}

Software failures can have enormous consequences in terms of
threatening peoples lives as well as economic loss because various critical systems
rely on software. Furthermore, software becomes increasingly
complex, which makes the prevention of failures even more
difficult. However, software quality assurance accounts already for
around 50\% of the development time \cite{myers04old}. Therefore it is
important to improve defect-detection techniques as well as reduce their
costs. Automation can be an option in that direction. For example,
automated test-case generation based on executable models is also 
under investigation as a possibility to make testing more
efficient \cite{pretschner05}.

Extensive research has been done on finding defects in code by automated
static analysis using tools 
called \emph{bug finding tools}, e.g.~\cite{flanagan02,ball02,hovemeyer04}. 
Although the topic is subject of ongoing investigations, there are only few
studies about how these tools relate among themselves and to other
established defect-detection techniques such as testing or reviews.

We will now discuss the problem situation in more detail. We briefly define
the terms we use in the following:
\emph{Failures} are a perceived
deviation of the output values from the expected values whereas \emph{faults}
are the cause of failures in code or other documents. Both are also referred
to as \emph{defects}. We use mainly \emph{defect} in our analyses also if
there are no failures involved as with defects related to maintenance only.

\paragraph{Problem.} We address the question of
how automated static analysis
using bug finding tools relates to other types of defect-detection techniques
and if it is thereby possible to reduce the effort for defect-detection using such tools.
In detail, this amounts to three questions.

\begin{enumerate}
\item Which types and classes of defects are found by different techniques?
\item Is there any overlap of the found defects?
\item How large is the ratio of false positives from the tools?
\end{enumerate}

\paragraph{Results.} The main findings are summarised in the following.

\begin{enumerate}
\item Bug finding tools detect a subset of the defect types that can be found
      by a review.
\item The types of defects that can be found by the tools can be analysed more
      thoroughly, that is, the tools are better regarding the bug patterns they
      are programmed for.
\item Dynamic tests find completely different defects than bug finding tools.
\item Bug finding tools have a significant ratio of false positives.
\item The bug finding tools show very different results in different
      projects.
\end{enumerate}

\paragraph{Consequences.} The results have four major implications.

\begin{enumerate}
\item Dynamic tests or reviews cannot be substituted by bug finding tools
      because they find significantly more and different defect types.
\item Bug finding tools can be a good pre-stage to a review because some
      of the
      defects do not have to be manually detected. A possibility
      would be to mark problematic code so that it cannot be overlooked
      in the review.
\item Bug finding tools can only provide a significant reduction in the
      effort necessary for defect-detection if their false positives ratios 
      can be
      reduced. From our case studies, we find the current ratios to be 
      not yet
      completely acceptable.
\item The tools have to be more tolerant regarding the programming style
      and design to provide more uniform results in different projects.
\end{enumerate}

\paragraph{Experimental Setup.} Five industrial projects and one 
development project from a university environment were selected which
are either already in use or in the final testing phase. We evaluated
several bug finding tools and chose three representatives that were
usable for the distributed web systems under consideration. The bug finding
tools and dynamic tests were used on all projects. A review was only possible
for a single project. The warnings issued from the tools were analysed
with experienced developers to classify them as true and false positives.
All  defects that were found were classified regarding their severity and defect
types. The comparison was done based on this classification.

\paragraph{Contribution.} We are not aware of studies that compare the
defects found by bug finding tools with the defects found by other
techniques, in particular not of any based on several, mainly industrial,
projects. A main contribution is also a thorough analysis of the ratio
of false positives of the bug finding tools as this is a significant
factor in the usability of these tools.

\paragraph{Organisation.} Sec.~\ref{sec:bft} gives an overview of bug
finding tools in general (Sec.~\ref{sec:bft_general}) and the three 
tools that were used in the projects (Sec.~\ref{sec:bft_analysed}).
The projects are described in
Sec.~\ref{sec:projects} with general characteristics in 
Sec.~\ref{sec:projects_general} and specific descriptions in 
Sec.~\ref{sec:projects_analysed}. The approach for the comparison 
of the techniques 
can be
found in Sec.~\ref{sec:approach} with a general discussion in
Sec.~\ref{sec:general}, the defect classification in
Sec.~\ref{sec:classification}, and the introduction of the defect types
in Sec.~\ref{sec:types}. The analysis of the study is described in
Sec.~\ref{sec:analysis} with the comparison among the bug finding tools
in Sec.~\ref{sec:comparison_bft}, bug finding tools versus reviews in
Sec.~\ref{sec:bft_review}, bug finding tools versus testing in
Sec.~\ref{sec:bft_testing}, and the defect removal efficiensies in
Sec.~\ref{sec:removal_efficiency}. We discuss our findings in 
Sec.~\ref{sec:discussion} and describe related work in Sec.~\ref{sec:related}. 
Finally, we conclude in Sec.~\ref{sec:conclusion} and sketch intended
future work in Sec.~\ref{sec:future}.

\section{Bug Finding Tools}
\label{sec:bft}

This section provides an introduction to bug finding tools in general
and describes briefly the three tools that were used in the case study.

\subsection{Basics}
\label{sec:bft_general}

Bug finding tools are a class of programs that aim to find 
defects in code by static analysis similarly to a compiler. The results 
of such a tool
are, however, not always real defects but can be seen as a warning that a
piece of code is critical in some way. There are various techniques to
identify such critical code pieces. The most common one is to define typical
bug patterns that are derived from experience and published common pitfalls
in  a certain programming language. Furthermore, coding guidelines and standards
can be checked to allow a better readability. Also, more sophisticated
analysis techniques based on the dataflow and controlflow are used. Finally, 
additional annotations in the code
are introduced by some tools \cite{flanagan02} to allow an extended static
checking and a combination with model checking.

\subsection{Analysed Tools}
\label{sec:bft_analysed}

The three bug finding tools that we used for the comparison are
described in the following. We only take tools into account that analyse
Java programs because the  projects we investigated, as described below, are all written
in that language. All three tools are published under an open source
license. We used these three tools as representatives for tools that mainly
use bug patterns, coding standards, and dataflow analysis, respectively. We deliberately
ignored tools that need annotations in the code because they have quite
different characteristics.

\paragraph{FindBugs.} The tool \emph{FindBugs} was developed at the
University of Maryland and can detect potentially problematic
code fragments by using a list of bug patterns. It can find faults
such as dereferencing null-pointers or unused variables. To some extent,
it also uses dataflow analysis for this. It analyses the software
using the bytecode in contrast to the tools described in the following.
 The tool is
described in detail in \cite{hovemeyer04}. We used the Version $0.8.1$
in our study.

\paragraph{PMD.} This tool \cite{pmd} concentrates on the source code and is
therefore especially suitable to enforce coding standards. It finds, for
example, empty try/catch blocks, overly complex expressions, and classes
with high cyclomatic complexity. It can be customized by using XPath
expressions on the parser tree. The version $1.8$ was used.

\paragraph{QJ Pro.} The third tool used is described in \cite{qjpro} and
analyses also the source code. It supports over 200 rules including 
ignored return values, too long variable names, or a disproportion between
code and commentary lines. It is also possible to define additional rules.
Furthermore, checks based on code metrics can be used. 
The possibility to use various filters is
especially helpful in this tool. We evaluated version $2.1$ in this study.

\section{Projects}
\label{sec:projects}

We want to give a quick overview of the five projects we analysed to
evaluate and compare bug finding tools with other defect-detection
techniques.

\subsection{General}
\label{sec:projects_general}

All but one of the projects chosen are  development projects from the
telecommunications company \O2\ Germany for backend systems 
with various development
efforts and sizes. One project was done by
students at the Technische Universit\"at M\"unchen. All these projects
have in common that they were developed using the Java programming
language and have an interface to a relational database system. The \O2\ projects
furthermore can be classified as web information systems as they all
use HTML and web browsers as their user interface.

\subsection{Analysed Projects}
\label{sec:projects_analysed}

The projects are described in more detail in \cite{koller04}. For 
confidentiality reasons, we
use the symbolic names A through D for the industrial projects.

\paragraph{Project A.} This is an online shop that can be used by customers
to buy products and also make mobile phone contracts. It includes
complex workflows depending on the various options in such contracts.
The software has been in use for six months. It consists of 1066
Java classes that  consist of over 58 KLOC (kilo lines of code).

\paragraph{Project B.} The software allows the user to pay goods bought
over the Internet using a mobile phone. The payment is added to the mobile
bill. For this, the client sends the mobile number to the shop and
receives a transaction number (TAN) via short message service (SMS). 
This TAN is used to authenticate the user
and authorises the shop to bill the user. The software has not been
put into operation at the time of the study. Software B has
215 Java classes with over 24 KLOC in total.

\paragraph{Project C.} This is a web-based frontend for managing
a system that is used to convert protocol files between different
formats. The
tool analysed only interacts with a database that holds administration
information for that system. The software was three months in use at the
time it was analysed. It consists of over 3 KLOC Java and JSP code.

\paragraph{Project D.} The client data of \O2\ is managed in the system
we call \emph{D}. It is a J2EE application with 572 classes, over
34 KLOC and
interfaces to various other systems of \O2.

\paragraph{EstA.} The only non-industrial software that we used in
this case study is \emph{EstA}. It is an editor for structuring
textual requirements developed during a practical course at the
Technische Universit\"{a}t M\"{u}nchen. It is a Java-based software
using a relational database management system. The tool has not been extensively
used so far. It has 28 Java classes with over 4 KLOC.

\section{Approach}
\label{sec:approach}

In this section, the approach of the case study is described. We start
with the general description and explain the defect classification and
defect types that are used in the analysis.

\subsection{General}
\label{sec:general}

We use the software of the five projects introduced in Sec.~\ref{sec:projects}
to analyse the interrelations between the defects found by bug finding
tools, reviews, and tests. For this, we applied each of these techniques
to each software as far as possible. While a review was only
made on project C, black-box as well as white-box tests were done
on all projects. We ran the bug finding tools with special care to be able
to compare the tools as well. To have a better possibility for comparison
with the other techniques, we also checked each warning from the bug finding
tools if it is a real defect in the code or not. This was done by an
inspection of the corresponding code parts together with experienced
developers. 
The usage of the techniques was completely
independent, that is, the testing and the review was not guided by results
from the bug finding tools. 

The external validity is limited in this case study. Although we 
mostly considered commercially developed software that is  in actual use, we only
analysed five systems. For better results more experiments are necessary.
Furthermore, the tests on the more mature systems, i.e.~the ones that are
already in use, did not reveal many faults. This can also limit the validity.
Moreover, the data from only one review is not representative but can only
give a first indication. Finally, we only analysed three bug finding tools, and
these are still under development. The results might be different if
other tools would have been used.

In the following we call all the warnings that are generated by the bug
finding tools \emph{positives}. \emph{True positives} are warnings that
are actually confirmed as defects in the code, \emph{false positives}
are wrong identifications of problems.

\subsection{Defect Categorisation}
\label{sec:classification}

For the comparison, we use a five step categorisation of the defects
using their severity. Hence, the categorisation is based on the
effects of the defects rather than their cause or type of occurrence
in the code. We use a standard categorisation for severity that is
slightly adapted to the defects found in the projects.
Defects in category 1 are the severest,
the ones in category 5 have the lowest severity. The categories are:

\begin{enumerate}
\item \emph{Defects that lead to a crash of the application.} These
      are the most severe defects that stop the whole application
      from reacting to any user input.
\item \emph{Defects that cause a logical failure.} This category consists of
       all defects that cause a logical failure of the application
      but do not crash it, for example a wrong result value.
\item \emph{Defects with insufficient error handling.} Defects in this
      category are only minor and do not crash the application or
      result in logical failures, but are not handled properly.
\item \emph{Defects that violate the principles of structured programming.}
      These are defects that normally do not impact the software but
      could result in performance bottlenecks etc.
\item \emph{Defects that reduce the maintainability of the code.} This
      category contains all defects that only affect the readability or
      changeability of the software.
\end{enumerate}

This classification helps us (1) to compare the various defect-detection
techniques based on the severity of the defects they find and (2) analyse
the types of defects that they find.

\subsection{Defect Types}
\label{sec:types}

Additionally to the defect classification we use defect types. That
means that the same or very similar defects are grouped together for an
easier analysis. This is not based on any standard types such as
\cite{beizer90,chillarege96,ieee93} but was defined specifically for the 
applications.

The defect types that we use for the bug finding tools can be seen
as a unification of the warning types that the tools are able to
generate. Examples for defect types are ``Stream is not closed''
or ``Input is not checked for special characters''.

\section{Analysis}
\label{sec:analysis}

This section presents the results of the case study and possible
interpretations. At first, the bug finding tools are compared among
each other, then the tools are compared with reviews, and finally with
dynamic tests.

\subsection{Bug Finding Tools}
\label{sec:comparison_bft}

We want to start with comparing the three bug finding tools described in 
Sec.~\ref{sec:bft} among themselves. The tools were used with each
system described above.

\subsubsection{Data.} Tab.~\ref{tab:bft} shows the defect
types with their categories and the corresponding positives found by each
tool over all  systems analysed. The number before the slash denotes
the number of true positives, the number after the slash the number of
all positives.

\begin{table}
\caption{Summary of the defect types found by the bug finding tools}
\begin{center}
\begin{tabular}{l r r r r}
\hline
Defect Type & Category & FindBugs & PMD & QJ Pro\\
\hline
Database connection is not closed   & 1 &   8/54 &   8/8 &   0/0\\
Return value of function ignored    & 2 &   4/4 &   0/0 &   4/693\\
Exception caught but not handled    & 3 &   4/45 &  29/217 &  30/212\\
Null-pointer exception not handled  & 3 &   8/108 &   0/0 &   0/0\\
Returning null instead of array     & 3 &   2/2 &   0/0 &   0/0\\
Stream is not closed                & 4 &  12/13 &   0/0 &   0/0\\
Concatenating string with + in loop & 4 &  20/20 &   0/0 &   0/0\\
Used ``=='' instead of ``equals''   & 4 &   0/1  &   0/0 &   0/29\\
Variable initialised but not read   & 5 & 103/103 &   0/0 &   0/0\\
Variable initialised but not used   & 5 &   7/7 & 152/152 &   0/0\\
Needless if-clause                  & 5 &   0/0 &  16/16 &   0/0\\
Multiple functions with same name   & 5 &  22/22 &   0/0 &   0/0\\
Needless semicolon                  & 5 &   0/0 &  10/10 &   0/0\\
Local variable not used             & 5 &   0/0 & 144/144 &   0/0\\
Parameter not used                  & 5 &   0/0 &  32/32 &   0/0\\
Private method not used             & 5 &  17/17 &  17/17 &   0/0\\
Empty finally block                 & 5 &   0/0 &   1/1 &   0/0\\
Needless comparison with null       & 5 &   1/1 &   0/0 &   0/0\\
Uninitialised variable in constructor & 5 & 1/1 &   0/0 &   0/0\\
For- instead of simple while loop   & 5 &   0/0 &   2/2 &   0/0\\
\hline
\end{tabular}
\end{center}
\label{tab:bft}
\end{table}

\subsubsection{Observations and Interpretations.}

Most of the true positives can be assigned to the 
category \emph{Maintainability
of the code}. It is noticeable that the different tools predominantly
find different positives. Only a single defect type was found from all
tools, four types from two tools each.

\begin{figure}[h]
  \centering \includegraphics[width=\textwidth]{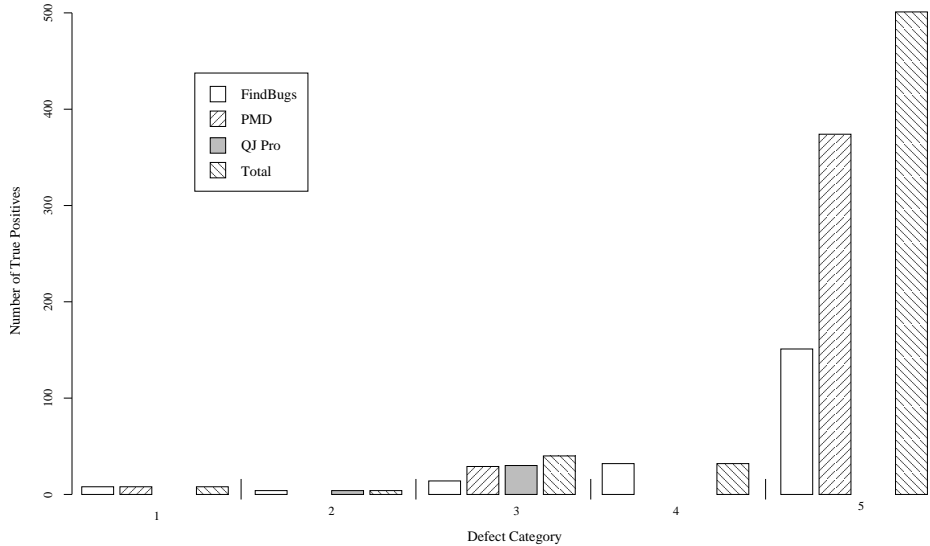}
  \caption{A graphical comparison of the number of true positives found by
           each tool and in total}
  \label{fig:graphic_stat}
\end{figure}

Considering the categories, FindBugs finds in the different systems
positives from all categories and
PMD only from the categories \emph{Failure of the application},
\emph{Insufficient error handling}, and \emph{Maintainability of the code}.
QJ Pro only reveals positives from the categories \emph{Logical failure of
the application}, \emph{Insufficient error handling}, and \emph{Violation
of structured programming}.
The number of faults found in
each category from each tool is graphically illustrated in
Fig.~\ref{fig:graphic_stat}.
Also the number of types of defects varies from tool to tool.
FindBugs detects defects of 13 different types, PMD of 10 types, and QJ Pro
only of 4 types.

The accuracy of the tools is also diverse. We use the defect type
``Exception is caught but not handled'' that can be found by all three tools
as an example. While FindBugs only finds 4 true positives, PMD reveals 29 and
QJ Pro even 30. For this, the result from QJ Pro contains the true
positives from
PMD which in turn contain the ones from FindBugs. A reason for this is
that QJ Pro is also able to recognize a single semicolon as an non-existent
error handling, whereas the other two interpret that as a proper handling.
This defect type is also representative in the way that FindBugs finds
the least true positives. This may be the case because it uses the compiled
class-files while PMD and QJ Pro analyse the source code.

A
further difference between the tools is the ratio of true positives
 to all positives.
PMD and FindBugs have a higher accuracy in
indicating real defects than QJ Pro. Tab.~\ref{tab:averages} lists the
average ratios of false positives for each tool and in total. It shows
that on average, half of the positives from FindBugs are false and still
nearly a third from PMD. QJ Pro has the worst result with only 4\% of
the positives being true positives. This leads to an overall average
ratio of $0.66$, which means that two thirds of the positives lead to
unnecessary work. However, we have to notice that FindBugs and PMD are
significantly better than that average.

\begin{table}
\caption{Average ratios of false positives for each tool and in total}
\begin{center}
\begin{tabular}{r r r r}
\hline
FindBugs & PMD & QJ Pro & Total\\
\hline
$0.47$ & $0.31$ & $0.96$ & $0.66$\\
\hline
\end{tabular}
\end{center}
\label{tab:averages}
\end{table}

An illustrative
example is the defect type ``Return value of function is ignored''.
FindBugs only shows 4 warnings that all are true positives, whereas QJ Pro
provides 689 further warnings that actually are not relevant. Because all
the warnings have to be looked at, FindBugs is in this case 
much more efficient than
the other two tools.

The efficiency of the tools varied over the projects. For the projects
\emph{B} and \emph{D}, the detection of the defect type
``Database connection not closed'' shows only warnings for true positives
with FindBugs.
For project \emph{A}, it issued 46 warnings for which the database
connection is actually closed. Similarly, the detection rate of true
positives decreases for the projects \emph{D} and \emph{A}
for the other two tools, with the exception of the well recognised
positives from the maintainability category by PMD. This suggests that
the efficiency of the defect detection depends on the design and the
individual programming style, i.e.~the implicit assumptions of the
tool developers about how ``good'' code has to look like.

A recommendation of usage of the tools is difficult because of the
issues described above. However, it suggests that QJ Pro, although
it finds sometimes more defects than the other tools, has the highest
noise ratio and therefore is the least efficient. FindBugs and PMD
should be used in combination because the former finds many different
defect types and the latter provides very accurate results in the 
maintainability category. Finally, PMD as well as QJ Pro can be used to enforce
internal coding standards, which was ignored in our analysis above.

\subsection{Bug Finding Tools vs.~Review}
\label{sec:bft_review}

An informal review was performed only  on project \emph{C}. The review team
consisted of three developers, including the author of the code. The reviewers
did not prepare specifically for the review but inspected the code
at the review meeting.

\subsubsection{Data.}

The review revealed 19 different types of defects which are summarised
in Tab.~\ref{tab:review} with their categories and number of occurrences.

\begin{table}
\caption{Summary of the defect types and defects found
         by the review}
\begin{center}
\begin{tabular}{l r r r r}
\hline
Defect Type & Category & Occurrences\\
\hline
Database connection is not closed & 1 & 1\\
Error message as return value & 1 & 12\\
Further logical case ignored & 2 & 1\\
Wrong result & 2 & 3\\
Incomplete data on error & 2 & 3\\
Wrong error handling & 2 & 6\\
ResultSet is not closed & 4 & 1\\
Statement is not closed & 4 & 1\\
Difficult error handling & 4 & 10\\
Database connection inside loop opened and closed& 4 & 1\\
String concatenated inside loop with ``+'' & 4 & 1\\
Unnecessary parameter on call & 5 & 51\\
Unnecessary parameter on return & 5 & 21\\
Complex for loop & 5 & 2\\
Array initialised from 1 & 5 & 21\\
Unnecessary if clauses & 5 & 8\\
Variable initialised but not used & 5 & 1\\
Complex variable increment & 5 & 1\\
Complex type conversion & 5 & 7\\
\hline
\end{tabular}
\end{center}
\label{tab:review}
\end{table}

\subsubsection{Observations and Interpretations.}

All defects found by
bug finding tools were also found by the review. However, the tools found
7 defects of type ``Variable initialised but not used'' in contrast to
one defect revealed by the review. On the other hand, the review detected
8 defects of type ``Unnecessary
if-clause'', whereas the tools only found one.
The cause is that only in the one defect that was found by both there was
no further computation after the if-clause. The redundancy of the others
could only be found out by investigating the logics of the program.

Apart from the two above, 17 additional types of defects were found, some
of which could have been found by tools. For example, the concatenation
of a string with ``+'' inside a loop is sometimes not shown by FindBugs
although it generally is able to detect this defect type.
Also, the defect that a database connection is not closed was not found,
because this was done in different functions. Furthermore it was not 
discovered by the tools
that the ResultSet and the corresponding Statement was never closed.

Other defect types such as logical faults or a wrong result from a function
cannot be detected by bug finding tools. These defects, however, can be
found during a review by following test cases through the code.

In summary, the review is more successful than bug finding tools, because it is
able to detect far more defect types. However, it seems to be beneficial
to first use a bug finding tool before inspecting the code, so that the
defects that are found by both are already removed. This is because the
automation makes it cheaper and more thorough than a manual review.
However, we also notice a high number of false positives from all tools.
This results in significant non-productive work for the developers that
could in some cases exceed the improvement achieved by the automation.

\subsection{Bug Finding Tools vs.~Testing}
\label{sec:bft_testing}

We used black box as well as white box tests for system testing the
software but also some unit tests were done. The black box tests were based
on the textual specifications and the experience of the tester. Standard
techniques such as equivalence and boundary testing were used. The white
box tests were developed using the source code and path testing. Overall
several hundred test cases were developed and executed. A coverage tool 
has also been used to check the quality of
the test suites. However, there were no stress tests which might
have changed the results significantly. Only for the projects \emph{EStA}
and \emph{C}, defects could be found. The other projects
are probably too mature to be able to find further defects by normal
system testing.

\subsubsection{Data.}

The detected defect types together with their
categories and the number of occurrences are summarised in 
Tab.~\ref{tab:testing}. We also give some information on the coverage data
that was reached by the tests. We measured class, method, and line
coverage. The coverage was high apart from project \emph{C}. In all the
other projects, class coverage was nearly 100\%, method coverage was also
in that area and line coverage lied between 60 and 93\%. The low coverage
values for project \emph{C} probably are because we invested the least
amount of effort in testing this project.

\begin{table}
\caption{Summary of the defect types and defects found
         by the tests}
\begin{center}
\begin{tabular}{l r r r r}
\hline
Defect Type & Category & Occurrences\\
\hline
Data range not checked & 1 & 9\\
Input not checked for special characters & 1 & 6\\
Logical error on deletion & 1 & 1\\
Consistency of input not checked & 2 & 3\\
Leading zeros are not ignored & 2 & 1\\
Incomplete deletion & 2 & 2\\
Incomprehensible error message & 3 & 7\\
Other logical errors & 2 & 3\\
\hline
\end{tabular}
\end{center}
\label{tab:testing}
\end{table}

\subsubsection{Observations and Interpretations.}

The defects found by testing are in the categories
\emph{Failure of the application}, \emph{Logical failure}, and
\emph{Insufficient error handling}.
The analysis above of the defects showed that the bug finding tools
predominantly find defects from the category \emph{Maintainability of
the code}. Therefore the dynamic test techniques find completely
different defects.

For the software systems for which defects were revealed, there were no 
identical defects
found with testing as well as the bug finding tools. Furthermore, the tools
revealed several defects also in the systems for which the tests were
not able to find one. These are defects that can only be found by extensive
stress tests, such as database connections that are not closed. 
This can only result
in performance problems or even a failure of the application, if the system
is under a high usage rate and there is a huge amount of 
database connections that are not closed. The most defects, however, are 
really concerning
maintainability and are therefore not detectable by dynamic testing.

In summary, the dynamic tests and the bug finding tools detect different
defects. Dynamic testing is good at finding logical defects that are
best visible when executing the software, bug finding tools have their
strength at finding defects related to maintainability. Therefore, we
again recommend using both techniques in a project.

\subsection{Defect Removal Efficiency}
\label{sec:removal_efficiency}

The defect removal efficiency is as proposed by Jones in \cite{jones91}
the fraction of all defects that were detected by a specific defect-detection
technique. The main problem with this metric is that the total number
of defects cannot be known. In our case study we use the sum of all different
defects
detected by all techniques under consideration as an estimate for this
number. The results are shown in Tab.~\ref{tab:efficiency}. The metric
suggests that the tools are the most efficient techniques whereas the
tests where the least efficient.

\begin{table}
\caption{The defect removal efficiencies per defect-detection technique}
\begin{center}
\begin{tabular}{l r r}
\hline
Technique & Number of Defects & Efficiency\\
\hline
Bug Finding Tools  & 585 & 76\%\\
Review             & 152 & 20\%\\
Tests              & 32 & 4\%\\
\hline
Total              & 769 & 100\%\\
\hline
\end{tabular}
\end{center}
\label{tab:efficiency}
\end{table}

However, we also have to take the defect categorisation into account
because this changes the picture significantly. The 
Tab.~\ref{tab:efficiency_category} shows the efficiencies for each
techniques and cateogory with the number of defects in brackets. It
makes obvious that tests and reviews are far more efficient in finding
defects of the categories 1 and 2 than the bug finding tools which
are the most severe defects.

\begin{table}
\caption{The defect removal efficiencies for each category}
\begin{center}
\begin{tabular}{r r r r r}
\hline
Category &  Bug Finding Tools & Reviews & Tests & Total\\
\hline
1  & 22\% (8) & 35\% (13) & 43\% (16) & 100\% (37)\\
2  & 15\% (4) & 50\% (13) & 35\% (9) & 100\% (26)\\
3  & 85\% (40) & 0\% (0) &  15\% (7) & 100\% (47)\\
4  & 70\% (32) & 30\% (14) & 0\% (0) & 100\% (46)\\
5  & 82\% (501) & 18\% (112) & 0\% (0) & 100\% (613)\\
\hline
\end{tabular}
\end{center}
\label{tab:efficiency_category}
\end{table}

\section{Discussion}
\label{sec:discussion}

The result that bug finding tools mainly detect defects that are
related to the maintainability of the code complies with the expectation
an experienced developer would have. Static analysis only allows to
look for certain patterns in the code and simple dataflow and controlflow
properties. Therefore only reviews or tests are able to verify the
logic of the software (as long as the static analysis is not linked
with model checking techniques). The tools do not ``understand'' the
code in that sense. The prime example for this is the varying efficiency
over the projects. In many cases, the tools were not capable to realise
that certain database connections are not closed in the same Java method
but a different one. They only search for a certain pattern. Therefore,
the limitation of static analysis tools lies in what is expressible by
bug patterns, or in how good and generic the patterns can be.

However, it still is surprising that there is not a single overlapping
defect detected by bug finding tools and dynamic tests. On the positive
side, this implies that the two techniques are perfectly complementary
and can be used together with great benefit. The negative side is that
by using the automated static analysis techniques we considered, it may not be possible to reduce
costly testing efforts. That there is only little overlapping follows
from the observation above that the tools mainly find maintenance-related
defects. However, one would expect to see at least some defects that
the tests found also detected by the tools, especially concerning
dataflow and controlflow. The negative results in this study can be
explained with the fact that most of the  projects analysed are quite
mature, and some of them are already in operation. This resulted in only
a small number of defects that were found during testing which in turn
could be a reason for the lack of overlapping.

A rather disillusioning result is the high ratio of false positives that
are issued by the tools. The expected benefit of the automation using
such tools lies in the hope that less human intervention is necessary
to detect defects. However, as on average two thirds of the warnings
are false positives, the human effort could be even higher when using
bug finding tools because each warning has to be checked to decide
on the relevance of the warning. Nevertheless, there are significant
differences between the tools so that choosing the best combination of
tools could still pay off. 

Bug finding tools that use additional annotations in the code for
defect-detection could be beneficial considering the overlap of defects
with other techniques as well as the false positives ratio. The annotations
allow the tool to understand the code to a certain extent and therefore
permits some checks of the logic. This deeper knowledge of the code might
reduce the false positives ratio. However, to make the annotations requires
additional effort by the developers. It needs to be analysed if this effort
is lucrative.

The effort and corresponding costs of the determination of defects using
the tools (including checking the false positives) was not determined
in this study. This is however necessary to find out if the use of bug
finding tools is beneficial at all.

\section{Related Work}
\label{sec:related}

There are only few
studies about how bug finding tools  relate among themselves and to other
established defect-detection techniques such as testing or reviews.

In \cite{rutar04} among others PMD and FindBugs are compared based on
their warnings which were not all checked for false positives. The findings
are that although there is some overlap the warnings generated by the
tools are mostly distinct. We can support this result with our data.

Engler and Musuvathi discuss in \cite{engler04} the comparison of their
bug finding tool with model checking techniques. They argument that static
analysis is able to check larger amounts of code and find more defects
but model checking can check the implications of the code not just
properties that are on the surface.

In \cite{johnson04} a static analysis tools for C code is discussed. The
authors state that sophisticated analysis of, for example, pointers leads
to far less false positives than simple syntactical checks.

An interesting combination of static analysis tools and testing in described
in \cite{csallner05}. It is proposed to use static analysis to find
potential problems and automatically generate test cases to verify if there
is a real defect. However, the approach obviously does not work with
maintenance-related defects.

Bush et al.\ report in \cite{bush00} on a static analyser for C and C++ code which is 
able to find several more dynamic programming errors. However, a comparison 
with tests was not done. Nevertheless, our observation that the defect-finding 
capabilities depend strongly on the coding styles of different programmers is supported in this paper.

In \cite{zitser04}, an evaluation of static analysis tools for C code regarding buffer overflows is described. 
The defects were injected and the fraction of buffer overflows found by each technique was measured. 
It is also noted that the rates of false positives or false alarms are unacceptably high.

Palsberg describes in \cite{palsberg01} some bug finding tools that use type-based analysis. He shows that they 
are able to detect race conditions or memory leaks in programs.

\section{Conclusion}
\label{sec:conclusion}

The work presented is not a comprehensive empirical study but a case study using
a series of projects
mainly from an industrial environment giving first indications of
how the defects found by bug finding tools relate to other defect-detection
techniques.

The main findings are that the bug finding tools revealed completely
different defects than the dynamic tests but a subset of the types of
the review. The defect types that are detected by the tools are analysed
more thoroughly than with reviews. The effectiveness of the tools seems
to strongly depend on the personal programming style and the design
of the software as the results differed strongly from project to project.
Finally, a combination of the usage of bug finding tools together with
reviews and tests would be most advisable if the number of false positives
were lower. It probably costs more time to resolve the false positives
than is saved by the automation using the tools.

Therefore, the main conclusion is that bug finding tools can save costs
when used together with other defect-detection techniques, if the tool
developers are able to improve the tools in terms of the false positives
ratio and tolerance of different programming styles.

\section{Future Work}
\label{sec:future}

This study is only a first indication and needs further empirical
validation to be able to derive solid conclusions. For this, we plan
to repeat this study on different subjects and also taking other tools
into account, e.g.~commercial tools or tools that use additional annotations
in the source code. Also, the investigation of other types of software
is important, since we only considered web applications in this study.

How the proper combination
of the different techniques can be found is also subject to further research.
As a first step more reliability-oriented measures, such as the 
failure intensity
efficiency \cite{wagner04a,wagner04b} can be used to 
compare the bug finding tools with other techniques. This can give more
clues in terms of the effect on the reliability of the usage of bug finding
tools. However, a comprehensive treatment of the subject needs to 
incorporate the false positives ratio into a cost model
based on \cite{wagner05d} to be able to determine the economically best
alternatives.

\subsection*{Acknowledgments}

We want to thank the authors of the tools FindBugs, PMD, QJ Pro for
investing such an amount of work in the tools and making them available
to the public.

\bibliographystyle{plain}
\bibliography{bibliography}

\end{document}